\documentclass[twocolumn]{aastex62}
\usepackage{amsmath,amstext}
\usepackage{tablefootnote}

\shorttitle{The radio emission of J1115+0544}

\newcommand{\erg}{${\rm erg \ s^{-1}}$ }

\def\ltsima{$\; \buildrel < \over \sim \;$}
\def\simlt{\lower.5ex\hbox{\ltsima}}
\def\gtsima{$\; \buildrel > \over \sim \;$}
\def\simgt{\lower.5ex\hbox{\gtsima}}
\newcommand{\msun}{{\rm\,M$_\odot$}}

\newcommand{\swift}{{\it\,Swift }}

\newcommand{\src}{{\rm\,J1115+0544 }}
\newcommand{\srcs}{{\rm\,J1115+0544}}

\begin{document}

\title{
Discovery of a years-delayed radio flare from an unusually slow-evolved tidal disruption event
}
\correspondingauthor{Xinwen~Shu} 
\email{xwshu@ahnu.edu.cn}

\author{Zhumao Zhang}
\affil{Department of Physics, Anhui Normal University, Wuhu, Anhui 241002, China}

\author[0000-0002-7020-4290]{Xinwen~Shu}
\affil{Department of Physics, Anhui Normal University, Wuhu, Anhui 241002, China}

\author{Lei Yang}
\affil{Department of Physics, Anhui Normal University, Wuhu, Anhui 241002, China}

\author{Luming Sun}
\affil{Department of Physics, Anhui Normal University, Wuhu, Anhui 241002, China} 

\author{Hucheng~Ding }
\affil{Department of Physics, Anhui Normal University, Wuhu, Anhui 241002, China} 

\author[0000-0003-1710-9339]{Lin Yan}
\affil{Caltech Optical Observatories, California Institute of Technology, Pasadena, CA 91125, USA}

\author{Ning Jiang}
\affil{Department of Astronomy, University of Science and Technology of China, Hefei, Anhui 230026, China}

\author{Fangxia An}
\affil{Purple Mountain Observatory, Chinese Academy of Sciences,
10 Yuan Hua Road, Qixia District,
Nanjing 210023, China}
\affil{Inter-University Institute for Data Intensive Astronomy (IDIA), Department of Physics and Astronomy, University of the Western Cape, 7535 Bellville, Cape Town, South Africa}

\author{Walter Silima}
\affil{Inter-University Institute for Data Intensive Astronomy (IDIA), Department of Astronomy, University of Cape Town, 7701 Rondebosch, Cape Town, South Africa}

\author{Fabao~Zhang }
\affil{Department of Physics, Anhui Normal University, Wuhu, Anhui 241002, China} 

\author[0000-0002-3698-3294]{Yogesh Chandola}
\affiliation{Purple Mountain Observatory, Chinese Academy of Sciences,
10 Yuan Hua Road, Qixia District,
Nanjing 210023, China} 

\author{ Zhongzu~Wu}
\affil{College of Physics, Guizhou University, Guiyang 550025, China} 

\author{Daizhong~Liu}
\affiliation{Purple Mountain Observatory, Chinese Academy of Sciences,
10 Yuan Hua Road, Qixia District,
Nanjing 210023, China} 

\author{Liming~Dou}
\affil{Department of Astronomy, Guangzhou University, Guangzhou 510006, China} 

\author{Jianguo Wang}
\affil{Yunnan Observatories, Chinese Academy of Sciences, Kunming 650011, China}

\author{Yibo Wang}
\affil{Department of Astronomy, University of Science and Technology of China, Hefei, Anhui 230026, China}

\author{Chenwei Yang}
\affil{Polar Research Institute of China, 451 Jinqiao Road, Shanghai 200136, China}

\author{Di~Li}
\affil{Department of Astronomy, Tsinghua University, Beijing 100084, China}
\affil{National Astronomical Observatories, Chinese Academy of Sciences, Beijing 100101, China}
\affil{Zhejiang Lab, Hangzhou, Zhejiang 311121, China}

\author{Tianyao~Zhou }
\affil{Department of Physics, Anhui Normal University, Wuhu, Anhui 241002, China}

\author{Wenjie~Zhang}
\affil{National Astronomical Observatories, Chinese Academy of Sciences, Beijing 100101, China}

\author{Fangkun~Peng }
\affil{Department of Physics, Anhui Normal University, Wuhu, Anhui 241002, China}

\author{Tinggui Wang}
\affil{Department of Astronomy, University of Science and Technology of China, Hefei, Anhui 230026, China}

\begin{abstract}

SDSS J1115+0544 is a unique low-ionization nuclear emission-line region (LINER) galaxy with energetic ultraviolet (UV), optical and mid-infrared outbursts occurring in its nucleus. 
We present the results from an analysis of multi-wavelength photometric and radio follow-up observations covering a period 
of $\approx$9 years since its discovery. 
We find that following a luminosity plateau of $\approx$ 500 days, the UV/optical emission has decayed back to the pre-outburst level, suggesting that the nuclear outburst might be caused by a stellar tidal disruption event (TDE). 
{In this case, \src could be an unusually slow-evolved optical TDE with longest rise and decline time-scales ever found.} 
Three years later than the optical peak, a delayed radio brightening was found with a
luminosity as high as $\nu L_{\nu}(\rm {5.5}~GHz)$ $\sim 1.9 \times 10^{39}$ \erg.  
Using a standard equipartition analysis, 
we find the outflow powering the radio emission was launched at $t$$\approx$1150 days 
with a velocity of $\beta \simlt$0.1 and kinetic energy of $E_{\rm K}\simgt10^{50}$ erg. 
The delayed radio brightening coupled with the disappearing plateau in the UV/optical light curves 
is consistent with the scenario involving 
delayed ejection of an outflow from a state transition in the disk.  
SDSS J1115+0544 is the first TDE displaying both a short-lived UV/optical plateau emission and a late-time radio brightening. 
Future radio observations of these TDEs in the post-plateau decay phase 
will help to establish the connection between outflow launching and changes in accretion rate. 

\end{abstract}


\section{Introduction} \label{sec:intro}

%
A star approaching close enough to a supermassive black hole (SMBH) 
can be torn apart when the tidal force of the SMBH exceeds the star’s self-gravity \citep{Rees1988}. 
Approximately half of the stellar debris remains bound and eventually to be accreted \citep{Bonnerot2020}, 
producing panchromatic flares peaking mainly at optical/ultraviolet (UV) and soft X-rays. 
These tidal disruption events (TDEs) are not only signposts of otherwise dormant SMBHs, but also a unique laboratory {for studying} 
the onset of accretion, the complex dynamics of material in the presence of strong gravitational fields, as well as any accompanying outflows associated with the accretion onto the SMBH \citep[e.g.,][]{Guillochon2015, Metzger2016}. 


Currently optical time-domain surveys dominate the discovery of TDEs, and 
dozens of candidates have been found, allowing for the population studies such as host properties, volumetric rates, luminosity function, 
and the correlations between {light curve} properties and BH masses \citep{vanVelzen2021, Hammerstein2023, Lin2022, Yao2023}. 
While most optically-selected TDEs are found to peak 
at a luminosity of $\sim$$10^{43}-10^{44}$\erg, followed by a smooth power-law decay at a rate of $t^{-5/3}$, deviations also exist with a plateau or rebrightening in the post-peak luminosity evolution \citep[e.g.,][]{Wevers2019, Jiang2019, Malyali2021, Yao2023}.    
Possible reasons for the deviation of the light curve from the $t^{-5/3}$ law are the delayed accretion \citep{Wevers2019, Chen2022}, repeating partial TDEs \citep{Liu2023, Somalwar2023, Lin2024}, or even TDEs by SMBH binaries \citep{Shu2020, Wen2024}.  

While optical/UV and X-ray emission in the early and peak phase of TDE evolution is directly tracking the mass fallback and the onset of accretion, 
radio observations can reveal and characterize outflows from TDEs, including the launching of relativistic jets \citep[e.g.,][]{Zauderer2011, Andreoni2022}. 
 However, most optically-selected TDEs are radio silent within the first few months of discovery \citep[e.g.,][]{Alexander2020}. 
Recent observations suggest that there is increasing number of TDEs with late-time radio 
detection or re-brightening, typically 2-4 years after the discovery \citep{Horesh2021a, Cendes2022, Cendes2024, Zhang2024}. 
The nature of the late-time radio brightening is still under debate, 
and several scenarios have been proposed, 
including decelerating and expansion of an off-axis jet launched at the time of disruption \citep[e.g.,][]{Giannios2011, Matsumoto2023}, 
a delayed launching of the outflow perhaps from delayed disk formation \citep{Horesh2021a, Cendes2024}, 
changes in the structure and density profile of circumnuclear medium (CNM)\citep[e.g.,][]{Mou2021, Mou2022, Zhuang2024, Matsumoto2024},
and break out of the choked precessing jets from the wind envelop \citep{Teboul2023, Lu2024}. 


SDSS J111536.57+054449.7 (hereafter \srcs) was initially identified through a systematic search for mid-infrared (MIR) outbursts in nearby galaxies \citep[MIRONG, ][]{Jiang2021}, using the archival data from the Wide-field Infrared Survey Explorer (WISE). 
In January 2015, \src underwent a strong optical flare, resulting in a brightening of 2.5 mag (V-band) relative to the galaxy brightness over $\sim$120 days. 
\citet[][hereafter Yan19]{Yan2019} performed a comprehensive analysis of its multi-wavelength photometric and spectroscopic observations up to $\sim$1000 days after the optical peak\footnote{Hereafter, all the phases refer to the rest-frame days relative to the time of optical peak (MJD = 57170). }, and found 
a slowly evolving post-peak optical light curve and non-variant broad Balmer lines over a period of at least 350 days. 
These properties are unusual for a typical optically-selected TDE, leading Yan19   
to propose that the flare in \src is from a new ``turn-on” accretion disk around SMBH, transforming the previously quiescent state into a type-1 AGN. 
However, based on the multi-epoch spectroscopic follow-up observations spanning a longer timescale of $\sim$3 years, \citet{Wang2022} found that the broad Balmer lines in \src have almost disappeared since January 2021, indicating a short-lived accretion event likely due to a TDE. 

In this paper, we present the results from the analysis of the 
complete multi-wavelength dataset spanning 9 years since its discovery.  
We find that both the optical/UV and MIR emission have decayed back to the preflare level, consistent with the TDE scenario. 
In addition, 
we report the late-time detection of radio brightening in \srcs, which can be explained 
in the framework of 
an energetic ($E_{\rm K}\simgt10^{50} {\rm erg}$) 
outflow launched with a significant delay of $t$$\approx$1150 days after optical peak. 

\section{observation and data reduction} \label{sec:obs_data}

\subsection{Optical and UV data}

As shown in Figure \ref{fig:opt_lc} (left), we collected the optical light curves of \src obtained 
by the Catalina Real-time Transient Survey \citep[CRTS,][]{Drake2009} and 
the Asteroid Terrestrial Impact Last Alert System (ATLAS)\footnote{https://fallingstar-data.com/forcedphot/}. 
The CRTS data are taken from Yan19 in which a flattening 
in the light curve up to 2017 May was found. 
Unfortunately, the CRTS data observed with the new camera is not yet available. 
Since the ATLAS photometry in individual observations has larger errors, we 
binned the light curve with a time period of 180 days in $c$-band and 120 days in $o$-band 
to increase the signal to noise ratio (S/N). 
Interestingly, the ATLAS data revealed that following the plateau phase, the optical emission started to decrease around MJD=57900 days. 
We also examined the public optical light curves of \src from the Zwicky Transient Facility (ZTF)\footnote{https://ztf.snad.space/}, 
but the data do not cover the plateau phase.  
Since most of the ZTF observations were performed after the ending of the flare, we do not consider the ZTF data in the following analysis. 

Since its optical discovery, \src has been observed in the three UV filters (UVW2, UVM2, UVW1) by {\it Swift} UVOT 
(the Ultra-violet Optical Telescope) for 9 times, with the latest one taken on 2023 Dec. 
By giving the source coordinate from optical position, we obtained magnitudes in the three 
bands for each observation using the task \textit{uvotsource}. 
 We performed flux extraction for source and background using a circular region with 5\arcsec\ radius and 
 a source-free circular region with 20\arcsec\ radius, respectively. 
 All the UV magnitudes are {referred to using} the AB magnitude system. 
 Similar to the optical light curves, it can be seen that the UV emission has decayed, 
 though the sampling of the light curves is sparse. 

\subsection{X-ray data}

As in the UV bands, \src has also been observed in the X-rays by {\it Swift} XRT (X-ray Telescope) 
over the same period. 
  Some results from the \swift/XRT observations in 2017 were presented in Yan19, while the data after 2019 have not yet been 
  fully reported. 
 After reducing the data following standard procedures in \textit{xrtpipeline}, we used {\tt Heasoft} (v6.33) to extract the  
 spectrum with the task \textit{xselect}. The source spectra were uniformly extracted in a circular region with a 40\arcsec\ radius, 
 and we selected an annular region with an inner radius of 60\arcsec\ and an outer radius of 120\arcsec\ for the background. 
While \src is not detected in individual \swift/XRT observations, coadding all of the XRT data results in a marginal detection 
with a count rate of 9.59 $\times 10^{-4}$ cts ${\rm s^{-1}}$ in the 0.3--10 keV. 
This corresponds to a flux of 3.78 $\times 10^{-14}$ erg ${\rm s^{-1} cm^{-2}}$ and luminosity of 7.47 $\times 10^{41}$ erg ${\rm s^{-1}}$, assuming a spectral model consisting of a powerlaw($\Gamma$ = 1.7) with HI Column Density of 5.43 $\times 10^{20} {\rm cm^{-2}}$. 
By splitting the data into that in plateau and decay phase, i.e., stacking the data taken in 2017 and 2017--2023 separately, we found 
a count rate of 1.05 $\times 10^{-3}$ cts ${\rm s^{-1}}$  and 1.25 $\times 10^{-3}$ cts ${\rm s^{-1}}$, respectively. 
This suggests that \src is X-ray faint, and there is no evidence of transient brightening in the different 
accretion phases as traced by UV/optical emission.

\subsection{MIR data}

 We built MIR light curves by collecting photometric data at 3.4$\mu$m (W1) and 4.6$\mu$m (W2) from the 
 WISE survey up to 2023 Dec 11.  
 Details of the WISE photometry and light curve construction are given in \citet{Jiang2021}.  
The WISE light curves are displayed in the lower panel of Figure \ref{fig:opt_lc} (left). 
It is clear that the MIR radiation has continuously dropped after rising to peak, 
and returned to the post-flare level at t$\approx$2500 days since discovery.  
This is consistent with what is observed in the long-term optical and UV light curves, 
confirming that the flare event is short-lived.

\subsection{Radio data}

\begin{deluxetable}{ccccccccc}
\centering
\tablewidth{0pt}
\tablehead{
\colhead{Observatory} & \colhead{Date} & \colhead{Phase$^{\dag}$} & \colhead{$\nu$} & \colhead{$F_\nu$} \\
\colhead{} & \colhead{} & \colhead{(days)} & \colhead{(GHz)} & \colhead{(mJy/beam)}
}
\tablecaption{Radio observations of \src \label{tab:table1}}
\setlength{\tabcolsep}{3mm}{
\startdata
FIRST & 2000 Feb  & -5137 & 1.4 & $<1.010$ \\ 
VLA & 2018 Nov 17 & 1165 & 5.5 & $0.711\pm 0.037$ \\ 
 & 2020 Jun 25 & 1703 & 1.5 & $<0.310$ \\ 
 & 2020 Jun 26 & 1704 & 5.5 & $1.700\pm 0.086$ \\ 
 & 2021 Aug 06 & 2077 & 5.5 & $0.520\pm 0.075$ \\ 
 & 2021 Oct 19 & 2145 & 1.5 & $0.727\pm 0.042$ \\ 
 & 2021 Sep 30 & 2127 & 5.5 & $0.413\pm 0.021$ \\ 
 & 2021 Sep 24 & 2122 & 9.0 & $<0.155$ \\ 
 & 2021 Dec 29 & 2210 & 1.5 & $<0.740$ \\ 
 & 2022 Jan 26 & 2236 & 3.0 & $0.398\pm 0.023$ \\ 
 & 2022 Jan 24 & 2234 & 5.5 & $0.304\pm 0.016$ \\ 
VLASS & 2017 Nov 22 & 834 & 3.0 & $<0.703$ \\ 
 & 2020 Jul 21 & 1727 & 3.0 & $<0.800$ \\ 
 & 2023 Feb 16 & 2590 & 3.0 & $<0.855$ \\ 
VLBA & 2021 Mar 17 & 1946 & 8.4 & $0.406\pm 0.031$ \\ 
GMRT & 2019 Dec 20 & 1530 & 1.4 & $<0.545$ \\ 
 & 2021 Sep 03 & 2102 & 0.7 & $<0.907$ \\ 
 & 2021 Dec 15 & 2197 & 0.7 & $0.349\pm 0.055$ \\ 
 & 2021 Dec 19 & 2201 & 1.2 & $0.558\pm 0.058$ \\ 
MeerKAT & 2023 Oct 20 & 2816 & 0.8 & $0.285\pm 0.019$ \\ 
 & 2023 Oct 20 & 2816 & 1.3 & $0.311\pm 0.017$ \\ 
 & 2023 Oct 28 & 2823 & 2.2 & $0.205\pm 0.011$ \\ 
 & 2023 Nov 07 & 2833 & 3.1 & $0.190\pm 0.010$ \\  
\enddata}
\begin{flushleft}
$^{\dag}$ The phase refers to the rest-frame days relative to MJD = 57170.
\end{flushleft}
\end{deluxetable}

\subsubsection{VLA}

\src was not detected by Faint Images of the Radio Sky at twenty cm (FIRST) using the Karl G. Jansky Very Large Array (VLA), with a 5$\sigma$ upper limit on the peak flux of 1.01 mJy beam$^{-1}$.
As part of radio follow-up observations of the MIRONG sample \citep[e.g.,][]{Dai2020}, 
we used VLA to observe \src at C-band (centered at 5.5 GHz, project code: 18B-086), and a compact source was detected with a peak flux of 
$S_{\rm 5GHz}=0.71\pm0.01$ mJy. 
To further
study the origin of the radio emission, we initiated multiple VLA observing campaigns (project code: 20A-251; 21A-146; 21B-168) over a period of 3.2 years, covering a frequency range 1.5--9 GHz. 
The data were reduced using the Common Astronomy Software Applications (CASA, version 5.3.0) 
and the standard VLA data reduction pipeline (version 5.3.1). 
For the reduced data product, we inspected each spectral window and 
manually flagged channels affected by radio frequency interference (RFI).
The calibrated data were imaged using the {\tt CLEAN} algorithm with 
Briggs weighting and ROBUST parameter of 0, which helps to reduce 
side-lobes and achieve a good sensitivity. 
\src was clearly detected in all observations. 
We used the {\tt IMFIT} task in CASA to fit the radio emission component with a two-dimensional 
elliptical Gaussian model to determine the position, integrated and peak flux density. 
The radio emission at all bands is unresolved and no extended emission is detected. 
For consistency, only peak flux densities are used in our following analysis.  
The VLA observation log and flux density measurements are presented in Table 1.

We also searched for the radio emission at 3 GHz using
the archival data from the Very Large Array Sky Survey \citep[VLASS;][]{Lacy2020}, 
but found \src remains undetected over its three-epoch observations between 2017 Nov and 2023 Feb, 
with a 5$\sigma$ upper limit on the peak flux of $\sim$0.8 mJy beam$^{-1}$.


\subsubsection{VLBA}

We carried out Very Long Baseline Array (VLBA) observations at the location of \src 
on 2021 March 17 with its 10 antennas (project code: BS295). 
The observing frequency was centered at 8.4 GHz in the X band. The observation was 
performed in the phase-referencing mode to a nearby strong compact radio source. 
Phase-reference cycle times were 3.0 minutes, with 2.0 minutes {on-target} and 1.0 minutes for the phase calibrator. 
We also inserted several scans of the bright radio source 3C 273 for fringe and bandpass calibration with an integration time of 2.5 minutes 
for each scan. 
The resulting total on-source time is 6 hours. 
To achieve sufficiently high imaging sensitivity, we adopted the observational mode RDBE/DDC to 
use the largest recording rate of 2 Gbps, corresponding to a recording bandwidth of 256 MHz in each of the dual circular polarizations. 
We used the NRAO AIPS software to calibrate the amplitudes and phases of the visibility data, following the standard procedure 
from the AIPS Cookbook\footnote{\url{http://www.aips.nrao.edu/cook.html}}. 
The calibrated data were imported into the Caltech DIFMAP package \citep{Shepherd1997} for imaging and model-fitting. 

VLBA detects a compact source in the final cleaned image, which has a deconvolved size of 1.48 mas $\times$ 0.94 mas. 
The integrated and peak flux density for the source is 695 $\pm$ 59 $\mu$Jy and 406 $\pm$ 23 $\mu$Jy/beam, respectively. 
To further investigate whether the source is resolved or not, we used the task {\tt Modelfit} in DIFMAP to fit the radio emission component, but found no additional emission components in the residual map. 
Therefore, \src remains compact and unresolved at the resolution of VLBA observation, with an upper 
limit on its size of $<$ 1.58 pc. 

\subsubsection{uGMRT}

{\src was observed with the upgraded Giant Metrewave Radio Telescope (uGMRT) at band 5 
(with a central frequency at 1.2 or 1.4 GHz) on 2019 Dec 20, 2021 Sep 10, and 2021 Dec 15 (project code: 37$\_$133; ddtC206; 41$\_$065), 
and at band 4 (centered at 0.7 GHz) on 2021 Sep 3 and 2021 Dec 19. }
%
Flux calibration was conducted using 3C 147 and 3C 286, whereas the nearby source 1120+143 was also used to determine the 
complex gain solutions. 
We note that the band 5 observation of project ddtC206 cannot be imaged as the 
{short baselines have problems estimating gain solutions for the target.}
The data from the uGMRT observations were reduced using CASA (version 5.6.1) following standard procedures and by using a pipeline adapted from the CAsa Pipeline-cum-Toolkit for Upgraded Giant Metrewave Radio Telescope data REduction \citep[CAPTURE;][]{Kale2021}. We began our reduction by flagging known bad channels, and the remaining RFI was flagged with the {\tt flagdata} task using the clip and tfcrop modes. 
We then ran the task {\tt tclean} 
with the options of the {MS-MFS \citep[multi-scale multi-frequency synthesis,][]{Rau2011} deconvolver,} two Taylor terms (nterms=2), and W-Projection \citep{Cornwell2008} to accurately model the wide bandwidth and the non-coplanar field of view of uGMRT. 

{While \src was not detected in the first and second epoch uGMRT observations, it appeared in the third epoch 
as an unresolved source with a peak flux of {0.349 $\pm$ 0.055 mJy/beam  
at band 4 and 0.558 $\pm$ 0.058 mJy/beam at band 5.}
All the uGMRT flux density measurements are shown in Table 1.}


\subsubsection{MeerKAT}

We also conducted multi-frequency radio observations with the MeerKAT telescope (project code: SCI-20230907-XS-01) in the UHF, L, S0 and S4-band, centered at 0.8 GHz, 1.3 GHz, 2.2 GHz and 3.1 GHz, respectively. 
We used the `4K' wideband continuum mode to ensure a high sensitivity of $\sim$10 $\mu$Jy. 
The total time is about 1.9 hrs for each band, {of which 1.5 hrs were spent on source,} and 
0.4 hr on the flux and phase calibrators (J0408-6545 and J1058+0133). 
The MeerKAT data were reduced using the OxKAT software \citep{Heywood2020}, 
and the final images were cleaned with the the WSClean algorithm \citep{Offringa2017}. 
We then measured the
integrated and peak flux, following the same procedures
described above. 
Thanks to the high sensitivity of the MeerKAT observations, the source 
is detected in all four bands. 
The ratio of the integrated flux to the peak
flux is in the range 1.01–1.24, with a median value of 1.03,
suggesting that most, if not all, of the radio emission is
unresolved, consistent with observations from other telescopes, though the spatial resolution 
is very different. 


\begin{figure*}[htbp!]
\epsscale{1}
\plottwo{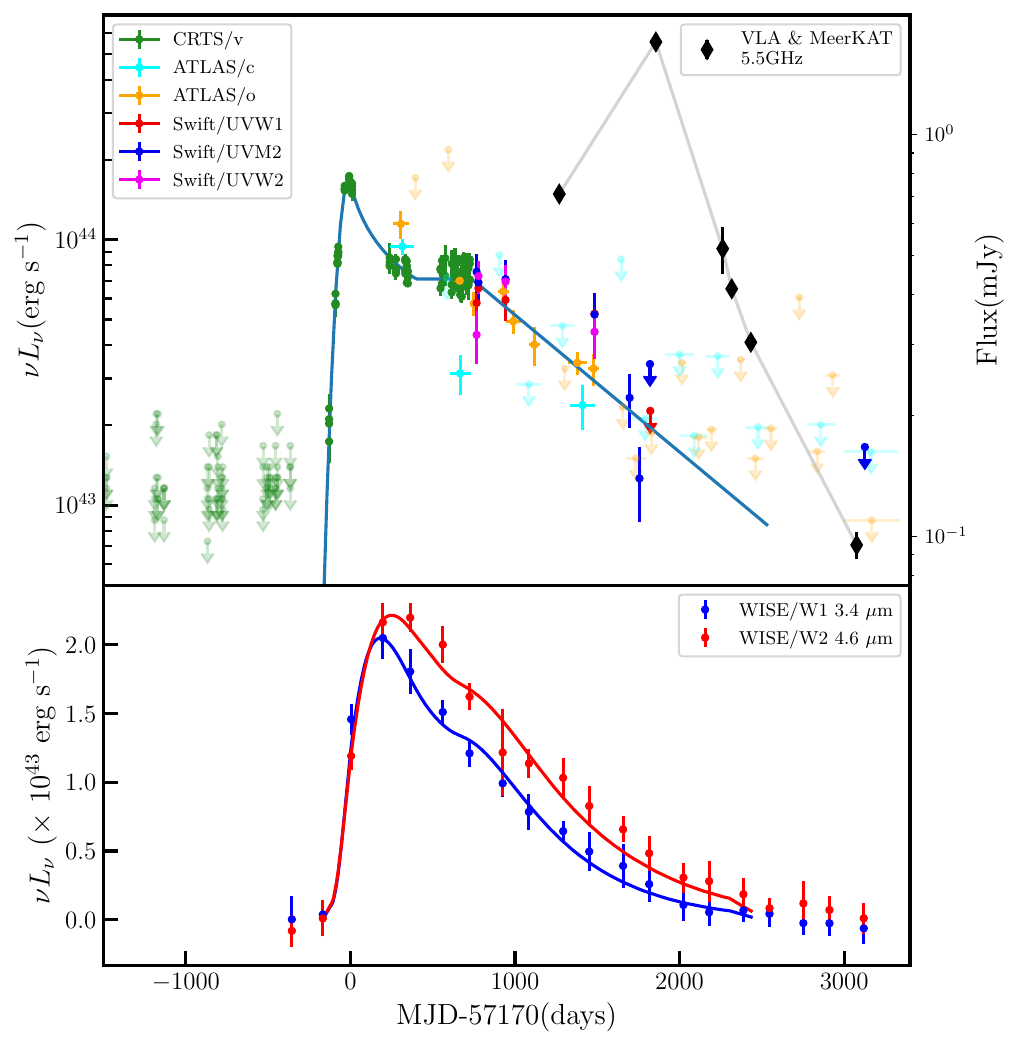}{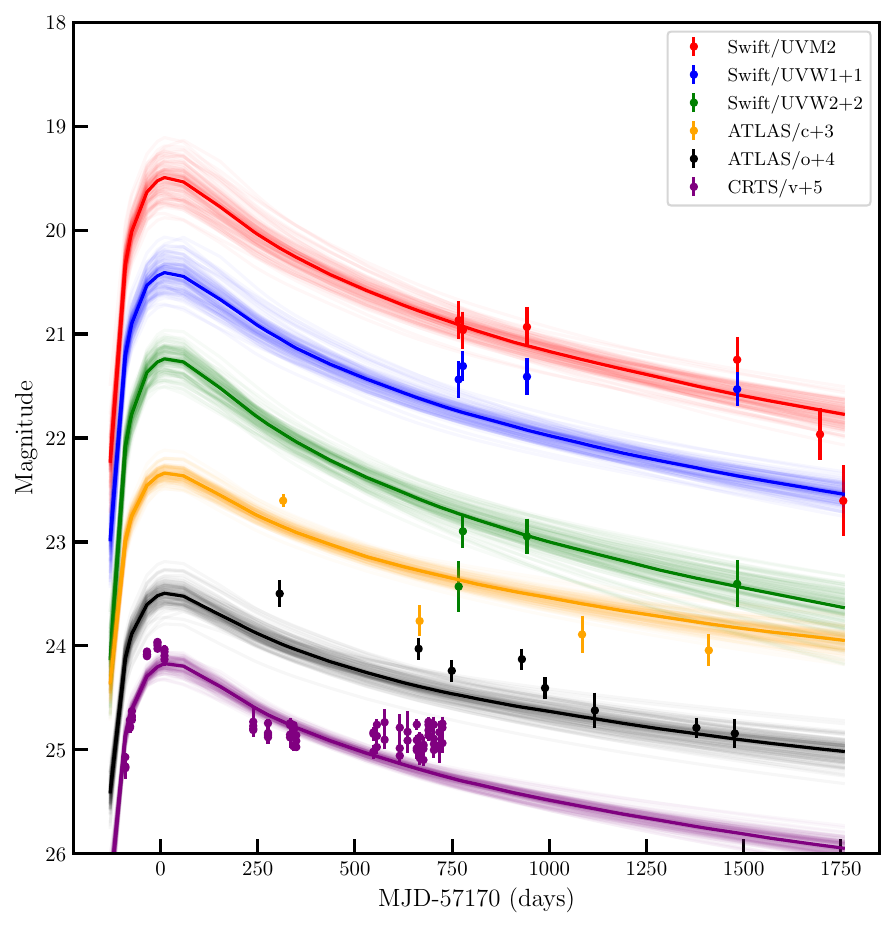}
\caption{The light curves of \src in different bands. 
Left panel: The upper panel shows the host-subtracted, dust extinction-corrected optical/UV light curves 
obtained from CRTS, ATLAS, and Swift/UVOT. 
{For comparison, we also show the radio light curve at 5.5 GHz, taken from the VLA and MeerKAT observations.  Since MeerKAT does not have the C-band receiver, we extrapolated the radio flux to 5.5 GHz based on the best-fit SED model (Figure 3)}. 
The blue solid line represent the four models to describe the optical/UV light curve in different evolution phases (Section 3.1).  
The lower panel shows the host-subtracted MIR light curves at 3.4$\mu$m and 4.6$\mu$m. 
The solid line represent the best-fit dust-echo models to account for the dust-reprocessed emission of central optical/UV flare, 
as detailed in the Section 3.1. 
Right panel: The optical/UV light curves with the best-fit model realizations from {\tt MOSFiT}. 
The photometry in different bands is shifted by constants for clarity. 
\label{fig:opt_lc}
}

\end{figure*}

\begin{figure}[htbp!]
\epsscale{1.15}
\plotone{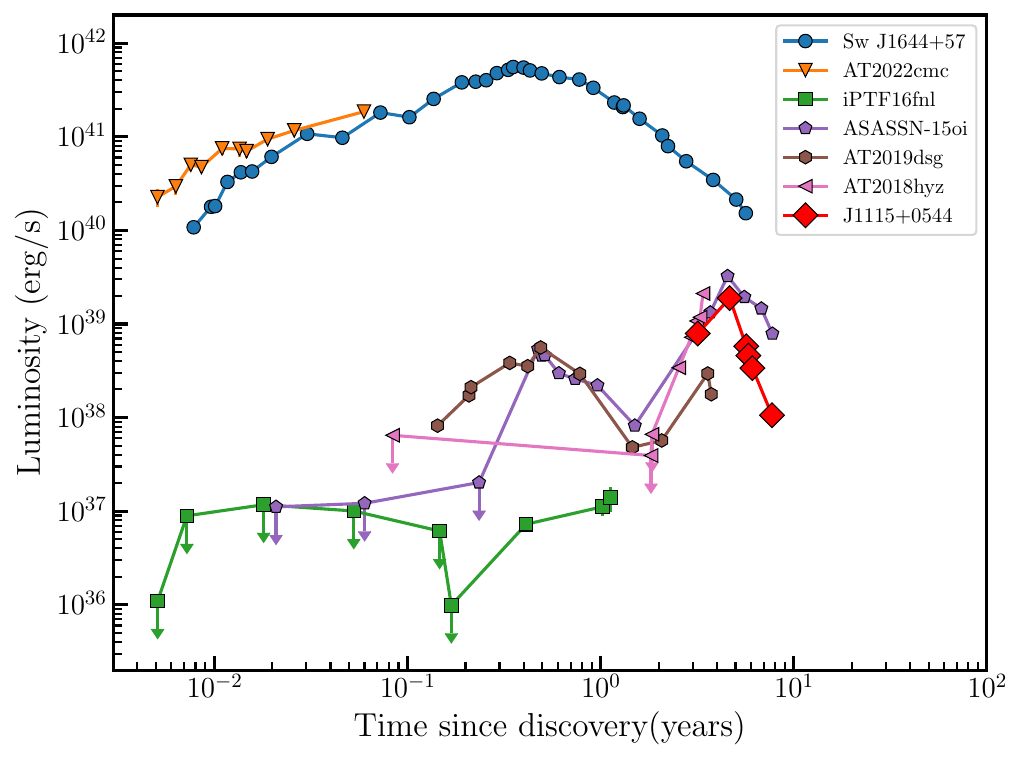}
\caption{\label{fig:radio_sed}
The radio luminosity evolution at 5.5 GHz for \src (red). 
Also shown for comparison are 
the light curves of the jetted TDEs Sw J1644+57 \citep{Berger2012,Zauderer2013,Eftekhari2018,Cendes2021b} 
and AT2022cmc \citep{Andreoni2022}, and other thermal TDEs with late-time radio brightening: iPTF16fnl from \citet{Horesh2021b}, ASASSN-15oi from \citet{Horesh2021a,Hajela2024}, AT2019dsg from \citet{Cendes2021a,Cendes2024}, 
and AT2018hyz from \citet{Cendes2022}.
\label{fig:radio_lc}
}
\end{figure}

\section{Analysis and results}

\subsection{Optical/UV and MIR Light-curve Analysis}

\src was previously considered as a turn-on AGN based on the appearance of post-flare plateau phase in the optical and MIR light curves \citep{Yan2019}, but only the photometric data up to $\sim$900 days post-flare were shown.  
By analyzing the new optical $o$- and $c$-band light curves of \src obtained by ATLAS, we found that the optical emission has 
faded back to the host level in 6 years (Figure \ref{fig:opt_lc}, left). 
Similar flux decline is found in the MIR light curves. 
This suggests that the plateau phase is a temporary behavior {if more data were collected.}

\par

To better characterize the bolometric luminosity evolution of the UV and optical emission, 
we fitted a model to the data by taking into account the effect of the dust extinction and assuming the the spectrum follows a blackbody with a constant temperature: 

$$
L(t,\nu) = 
\left\{
\begin{aligned}
& L_0 \exp\left(-\frac{(t-t_{\rm p})^2}{2\sigma^2}\right), & t<t_p \\
& L_0 \left(\frac{t-t_{\rm p}+\tau_1}{\tau_1}\right)^p, & t_p \leqslant t < t_f \\
& L_{\rm f} = L_0\left(\frac{t-t_{\rm f}+\tau_1}{\tau_1}\right)^p, & t_f \leqslant t < t_{\rm d} \\
& L_{\rm f} \exp\left(-\frac{t-t_{\rm d}}{\tau_2}\right), & t\geqslant t_d
\end{aligned}
\right.
$$
where
$$
L_0=L_{\rm peak}\frac{\pi B_\nu}{\sigma_{\rm SB} T^4}\times 10^{0.4 \Delta m_\nu(A_{\rm v})}
$$
{$B_\nu$ is the blackbody spectrum, $\sigma_{\rm SB}$ stands for the Stefan-Boltzmann constant, } 
{$T$ is the blackbody temperature, and $A_{\rm v}$ is the extinction magnitude in the v-band.}
The model describes the light curve by splitting it into four segments: 
the rising to peak followed by the power-law decay \citep{vanVelzen2021}, 
the plateau phase, and a rapid decline.
The meanings of the parameters are as follows: $L_{\rm peak}$ is the peak luminosity, $L_{\rm f}$ is the luminosity during the plateau phase. $t_{\rm p}$ is the time corresponding to the peak luminosity, $t_{\rm f}$ and $t_{\rm d}$ represents the time at which the plateau phase begins and ends, respectively. $\tau_1$ and $\tau_2$ are the decline timescales of the first and second decline stages, respectively. 
\par

We used a Markov chain Monte Carlo (MCMC) fitting technique \citep[python module {\tt emcee},][]{Foreman-Mackey2013} 
to determine the best-fitting parameters and uncertainties, which {were found to be: $\log L_{\rm peak}=44.23_{-0.41}^{+0.66}$ erg/s, $t_{\rm p}=-9.23_{-1.41}^{+1.94}$ days, $\sigma=56.81_{-1.23}^{+1.57}$ days, $\tau_1=72.21_{-22.25}^{+36.48}$ days, $p=-0.47_{-0.11}^{+0.07}$, $t_{\rm f}=396.62_{-17.46}^{+19.39}$ days, $t_{\rm d}=737.82_{-31.42}^{+87.02}$ days, $\tau_2=841.70_{-101.26}^{+85.86}$} days, $\log T=4.37_{-0.13}^{+0.22}$ K, $A_{\rm v}=0.88_{-0.32}^{+0.32}$. 
Note that the resulting blackbody temperature is consistent with that of the optical TDEs discovery by ZTF \citep{Hammerstein2023, Yao2023}. 
\par
Assuming that the MIR emission is originated from the dust echo of the observed UV-optical flare~\citep{Lu2016,Jiang2016,vV2016},
we tried to explore the dust properties by fitting the observed MIR light curves with {the dust radiative transfer model which describes the IR echo of a TDE} 
developed by \citet{Lu2016} and \citet{Sun2020}. 
There are four parameters related to the dust: inner radius $r_{\rm in}$, outer radius $r_{\rm out}$, dust grain size $a_0$, and dust density $n_{\rm d}$. 
We found that $r_{\rm in}$ is not well constrained with only an upper limit of $r_{\rm in}$ is $9.5 \times 10^{16} {\rm cm}$, $r_{\rm out} = 1.2_{-0.2}^{+0.2} \times 10^{18} {\rm cm}$, $a_0 \sim 1 {\rm \mu m}$ assuming a prior range of 0.01 to 1 ${\rm \mu m}$, and $n_{\rm d} = 6.6_{-2.4}^{+6.0} \times 10^{-11} {\rm cm^{-3}}$.
Note that the outer radius of dust distribution is comparable to that of the radio-emitting region inferred from an equipartition analysis (Section 3.3), suggesting that they are likely physically connected. 
We integrated the MIR blackbody luminosity and obtained that the energy released is $2.96 \times 10^{51}$erg, 
which is a factor of 3.5 lower the energy released in UV/optical bands. 
{
Since the bolometric output is dominated by the optical-UV emission, summing the optical-UV luminosity results in a total integrated energy of $1.02\times 10^{52}$ erg, which corresponds to an accreted mass of 0.06 \msun with an assumed accretion radiative efficiency of 0.1.} 
\par


We further performed joint fits to the optical and UV light curves using the TDE model \citep[{\tt MOSFiT},][]{Mockler2019}.  
In Figure \ref{fig:opt_lc} (right), we show an ensemble of model realizations from
{\tt MOSFiT}. 
The model is able to reasonably reproduce both the early- and late-time data, 
though it does underpredict the emission in the plateau phase of the CRTS v-band light curve. 
This is likely because the {\tt MOSFiT} TDE model does not incorporate the physical components
needed to fit the intermediate luminosity plateau, as it was built to account for UV/optical emission when the bolometric luminosity closely follows the fallback rate. 
The best-fit model invokes a SMBH ($\log (M_{BH}/M_\odot) = 7.50_{-0.17}^{+0.14}$) disrupting  
a star of $1.4_{-0.2}^{+0.3}$ $M_\odot$. 
{The black hole mass is consistent within uncertainties with the value estimated by \citet{Yan2019} based on the $M_{\rm BH}-\sigma_{\star}$ relation and BLR velocity dispersion. }
The best-fit model from {\tt MOSFiT} also implies that
the star was nearly fully disrupted by the black hole as the scaled impact parameter
$b=0.98_{-0.08}^{+0.07}$ \citep{Mockler2019}.  
{
It should be noted that the {\tt MOSFiT} models cannot describe the Swift/UVW1 data 
well, i.e., the observed magnitude is brighter than
we would expect from the {\tt MOSFiT} model shown in Figure \ref{fig:opt_lc} (right). 
This could be due to the significant temperature evolution of the UV-optical emission in \srcs, 
which cannot be effectively accounted for by {\tt MOSFiT}. 
In order to test the robustness of the fitting results, we excluded the Swift/UVW1 data from 
the {\tt MOSFiT} fittings, and found the relevant parameters are in good
agreement within {uncertainties} 
with the model of the full data set. 
}

\subsection{Radio Flux and Spectral Evolution}

The radio light curve of \src at 5.5 GHz (that is relatively well-sampled) is shown in Figure \ref{fig:opt_lc} (left panel). 
In comparison with the time of the optical discovery, 
we find a flux rise from about 0.7 mJy to 1.7 mJy over a period of 539 days, 
followed by a steep decline lasting for at least 1120 days. 
Fitting the light curve with a power-law ($F_{\rm \nu}\propto t^{\alpha}$), 
we find an index $\alpha=2.29$ and $\alpha=-6.36$ in the rising and declining phase, 
respectively. 

\begin{figure*}[ht!]
\epsscale{1.15}
\plottwo{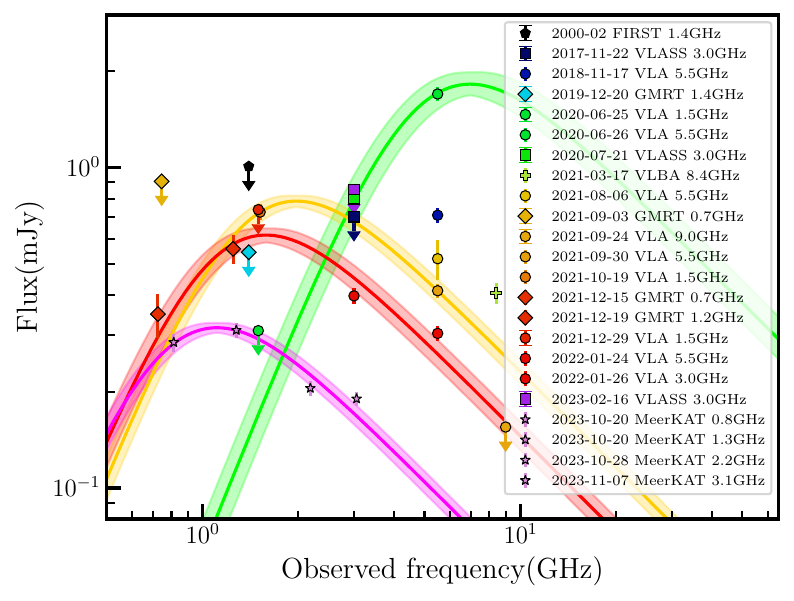}{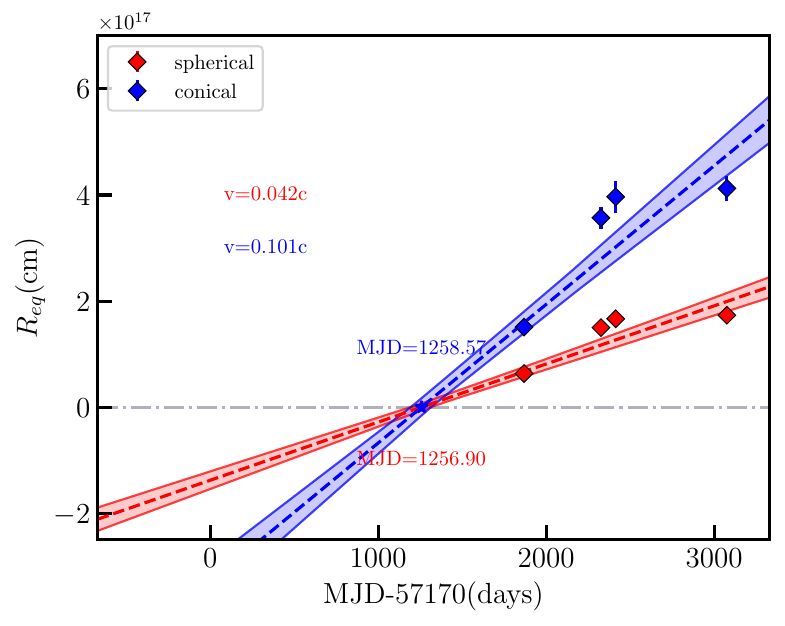}
\caption{
Left panel:
Radio SED and its evolution over four epochs, using the data from VLA, GMRT, VLBA and MeerKAT observations. 
For the nondetections, the corresponding 3$\sigma$ upper limits on flux density are shown.
The solid line represents the best-fit SED for each epoch, and the shaded region denotes the 1$\sigma$ error range 
of model realizations in MCMC fittings. 
Right panel:
The evolution of equipartition radius of the radio-emitting region as a function of time. 
Red and blue filled diamonds indicate parameters for a spherical homogeneous and a collimated
conical outflow, respectively.
We perform linear fittings to the data and obtain the free expansion velocity of the outflow for the two different geometries. 
Shaded regions represent the 1$\sigma$ range of the best-fit relation. 
Extrapolating the linear trend to the time at which $R_{\rm eq}=0$, we are able to constrain   
the outflow launching time relative to the time of optical discovery (MJD=57170). 
\label{fig:radio_req}
}
\end{figure*}

Figure \ref{fig:radio_lc} shows the radio luminosity evolution of \srcs. 
For comparison, we also plot the radio evolution of jetted TDEs 
and those with obvious late-time radio brightening on timescales of hundreds of days. 
The radio luminosity of \src increases from $7.90 \times$ 10$^{38}$ erg s$^{-1}$ at $t=1165$ days to $1.89 \times$ 10$^{39}$ erg s$^{-1}$ at $t=1704$ days. 
 This is similar 
 to the luminosity evolution of the TDE AT2018hyz \citep{Cendes2022} and secondary rising phase in the TDE ASASSN-15oi \citep{Horesh2021a, Hajela2024}. 
 Although the luminosity rise in \src appears shallower, its actual rise may be comparable to 
 AT2018hyz and ASASSN-15oi as only two data points were obtained with our radio follow-up observations. 
 The luminosity evolution of \src following its temporal peak is dominated by the optically thin emission, and 
 declining approximately linearly with time, which is steeper than that observed in ASASSN-15oi 
 and a predicted power-law index of $-3\simlt \alpha \simlt$$-1$ \citep{Generozov2017}. 
It should be noted that the radio luminosity of \src is one order of magnitude lower than that of the jetted TDE Sw J1644+57 at a comparable timescale ($\approx$1100-2000 days), 
ruling out the origin of radio brightening from a decelerated on-axis jet. 

Further insights into the nature of the radio emission can be obtained from the 
analysis of the radio spectral evolution as a function of time. 
Figure \ref{fig:radio_req} (left) shows the radio spectral energy distribution (SED) in the 
$\sim$$0.6-10$ GHz and its time evolution over a period of $\sim$1200 days. 
It is clear that the SED exhibits a gradual shift to a lower peak flux density and frequency. 
We model the SED evolution with the synchrotron emission models in the context of an outflow 
expanding into the CNM in which 
the blastwave 
amplifies the magnetic field and accelerates the
ambient electrons producing transient radio emission \citep[e.g.,][]{Zauderer2011, Alexander2016}. 
As shown in \citet{Goodwin2022, Cendes2022}, 
there are four parameters to characterize the synchrotron emission spectrum, including  
$F_{0}$, $\nu_{m}$, $\nu_{a}$, and $p$, where $F_{0}$ is the flux normalization at 
$\nu_{m}$ (the synchrotron minimum frequency), $\nu_{a}$ is the synchrotron self-absorption frequency, and $p$ 
is the energy index of the power-law distribution of relativistic electrons. 

Following the same approach outlined in \citet{Goodwin2022}, we use a 
MCMC fitting technique 
to perform the synchrotron model fits and determine the best-fitting parameters and uncertainties. 
The posterior distributions are sampled using 128 chains,
which were run for 4000 steps to ensure that the samples have sufficiently converged. 
Due to the limited data points 
we fix the power-law index of electrons' distribution  
to $p = 3$ \citep[e.g..][]{Alexander2016,Cendes2021b}. 
In Figure \ref{fig:radio_req} (left), we show the resulting SED models which provide a good fit to the data. 
From the SED fits we determine the peak flux density and frequency, $F_{\nu, p}$ and $\nu_{p}$, respectively.  
We find that both $F_{\nu, p}$ and $\nu_{p}$ indeed decrease steadily with time, from 1.82 mJy and 6.96 GHz at $t\sim 1716$ days to 0.32 mJy and 1.11 GHz at $t\sim 2825$ days. 

\subsection{Equipartition Analysis}
\begin{figure*}[ht!]
\epsscale{1.15}
\plottwo{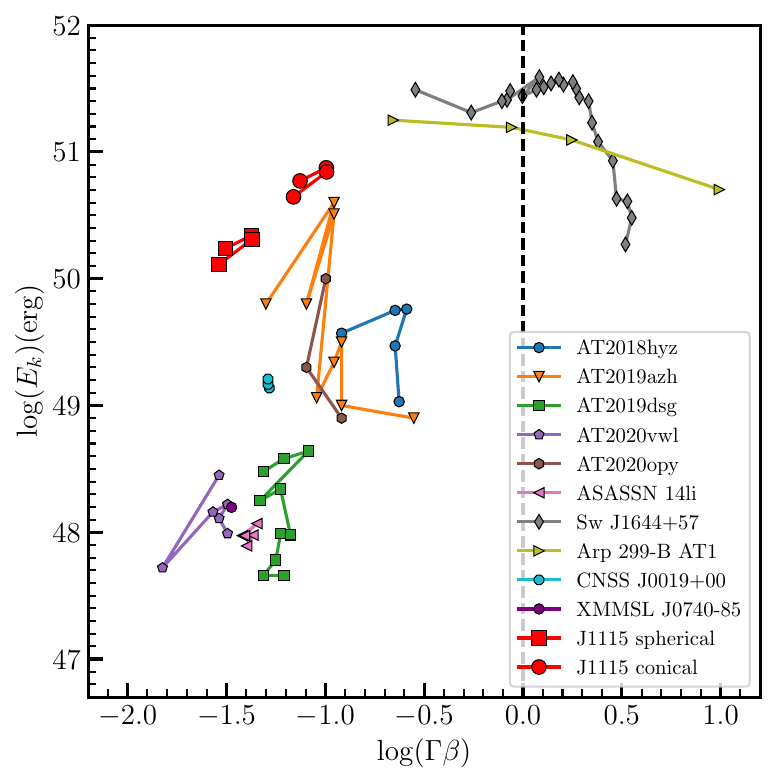}{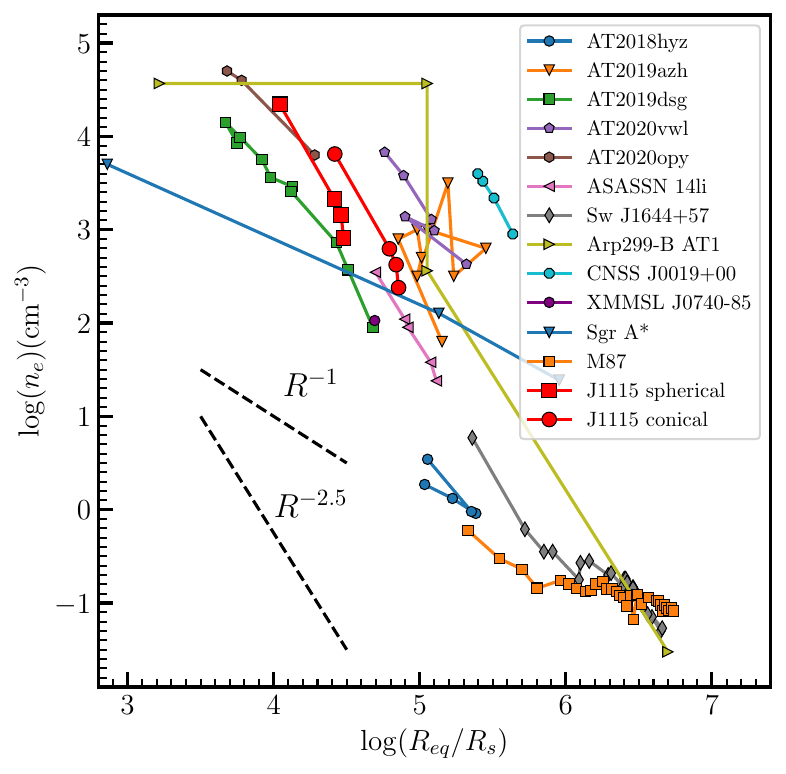}
\caption{Left Panel: Outflow velocity and kinetic energy of \src for the spherical case and collimated conical case.  
Vertical dashed line is shown to distinguish whether the outflow is relativistic or non-relativistic. 
For comparison, we include the jetted TDE Sw J1644+57 \citep{Berger2012,Zauderer2013,Eftekhari2018,Cendes2021b}, the TDE with an off-axis jet Arp 299-B AT1 \citep{Mattila2018}, 
and non-relativistic TDEs \citep{Alexander2016, Alexander2017, Anderson2020, Cendes2022, Goodwin2022, Cendes2021a, Goodwin2023a, Goodwin2023b}. 
Right Panel: CNM density profile for \srcs, normalized to the Schwarzschild radius of the SMBH with mass given by Yan19. 
We also show the radio-inferred density profiles for the TDEs shown on the left, and that measured using X-ray data for 
Sgr A* \citep{Gillessen2019} and M87 \citep{Russell2015}.   
\label{fig:param_compare}}
\end{figure*}

With the inferred values of $F_{\nu, p}$ and $\nu_{p}$, we can further use an equipartition analysis to derive 
the radius of the radio emitting region ($R_{\rm eq}$) and outflow kinetic energy ($E_{\rm eq}$)
using the scaling relations outlined in \citet{Barniol2013}. 
In order to account for possible geometric dependence of outflow evolution, we considered both a spherical outflow and a mildly collimated conical outflow with a half-opening angle of $\phi=30^{\circ}$ \citep[e.g.,][]{Goodwin2022}. 

In Figure \ref{fig:radio_req} (right), we show the evolution of $R_{\rm eq}$ assuming the two different outflow geometries. 
For the spherical outflow, we find that the radius increases slightly from $R_{\rm eq}\approx$ 6.34 $\times$ 10$^{16}$ cm  to $\approx$ 1.73 $\times$ 10$^{17}$ cm between $t=1716$ and 2825 days. 
The increase in $R_{\rm eq}$ becomes more rapidly for the case of a mildly collimated conical outflow. 
Under the assumption of free expansion, this corresponds to an outflow velocity {\bf ($\beta=v/c$)} of 0.05 and 0.11 for the spherical and conical geometry, 
respectively. 
The inferred outflow kinetic energy 
is $E_{\rm eq}>10^{50}$ erg (Figure 4, left),  
which appears to be larger than most other radio-emitting TDEs with non-relativistic outflows \citep{Cendes2024}. 
{Combining the outflow
velocity and kinetic energy\footnote{The kinetic energy of the radio-emitting region ($E_{\rm kin}$) and $E_{\rm eq}$ are not independent, and $E_{\rm kin}/E_{\rm eq}\sim\mathcal{O}(1)$ \citep{Matsumoto2022}.} we can infer an ejected mass of $M_{\rm ej}\approx0.07-0.11$\msun~dependent on the outflow geometry. 
Note that this is a lower limit on the mass, as in a realistic blast wave the energy of the radio-emitting region is the kinetic energy of the fraction of the outflow that was moving with a velocity larger than 
the current velocity \citep{Matsumoto2022}. 
This suggests that the mass of the disrupted star is larger than 0.2-0.3\msun, 
ruling out the tidal disruption of a lower-mass star of $\sim$0.1\msun~as found in most other optical TDEs \citep{Mockler2019}. 
}

The equipartition analysis suggests that the CNM density profile of \src is similar to that inferred for other non-relativistic TDEs (Figure 4, right). 
The inner portion of the density profile is approximately proportional to $R^{-2.5}$, 
while it shows evidence for a steepening at a radius of about $1.7\times10^{17}$ cm (2.9 $\times10^4$$R_{\rm S}$),   
similar to that observed in the TDE AT2019dsg \citep{Cendes2021a}, 
indicating that the CNM might have a stratified gas distribution. 

\section{Discussion} 
\subsection{\src revisited: an unusually slow-evolved TDE candidate in a LINER galaxy?}

\src was reported by \citet{Yan2019} as a ``turn-on" AGN, transitioning from a quiescent state to a type-1 AGN with a sub-Eddington accretion rate. 
In this case, the fast transitioning that occurred on a timescale of only $\sim$120--200 days,  
is difficult to reconcile with the conventional accretion disk theories. 
\citet{Yan2019} argued against \srcs's flare being a typical optically selected TDE, 
because the post-peak optical light curve settled into a plateau phase of at least 600 days following the initial decline of $\sim$200 days, 
with the simultaneous emergence of broad Balmer lines that remain non-variant for $\sim$400 days. 
However, based on the follow-up spectroscopy observations performed on 2021 Jan 7, $\sim$6 years since the discovery, \citet{Wang2022} 
found that its broad component of the H$\alpha$ emission line has almost disappeared. 
This is likely due to the decay in the continuum emission. 
By using the complete UV/optical and MIR light curves spanning 9 years since 
its discovery, we found unambiguous evidence that \srcs's flare has faded back to the 
baseline level in $\sim$5 yr (Figure \ref{fig:opt_lc}), and the post-peak plateau phase 
is short-lived (lasting for $\sim$490 days),
which is at odds with the persistent emission expected for a ``turn-on" AGN. 
Therefore, a TDE causing the multi-wavelength flares in \src seems more favored.




\begin{figure}[htbp!]
\epsscale{1.15}
\plotone{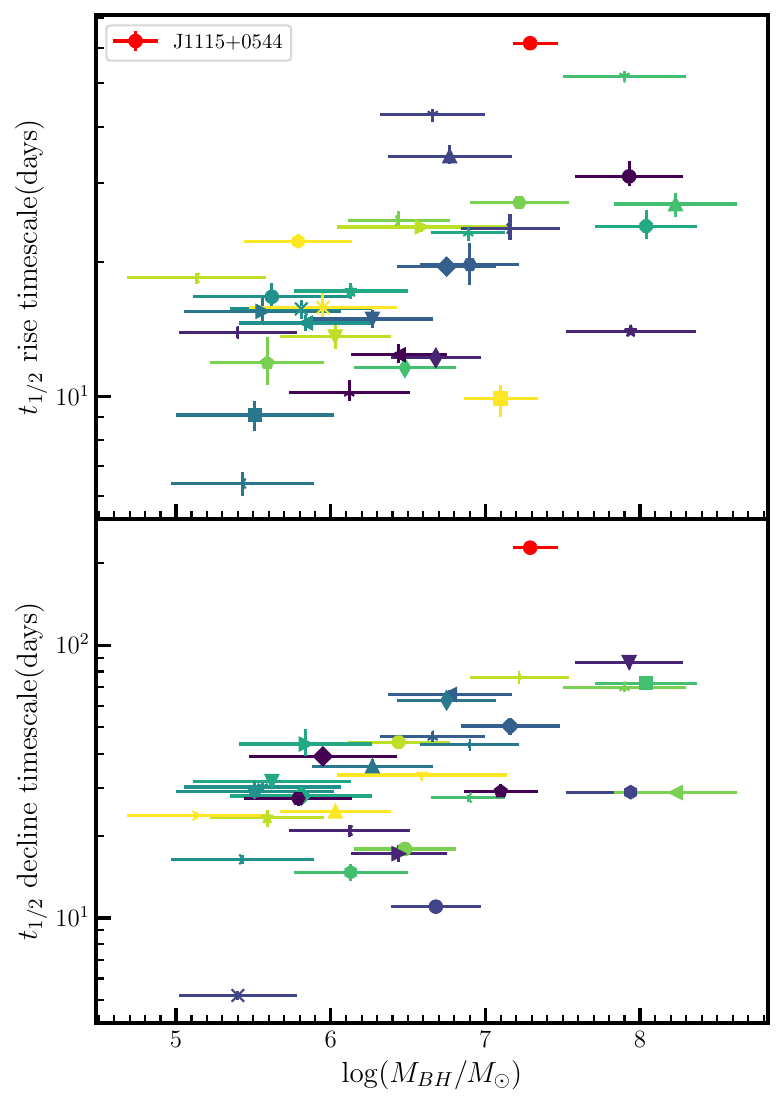}
\caption{
{The correlation between the black hole mass ($M_{\rm BH}$) and the rest-frame rise time-scales ($t_{\rm 1/2, rise}$, upper panel) and decline time-scales ($t_{\rm 1/2, decline}$, lower panel) for \src (red) and other optical TDEs \citep{Yao2023}. 
The $M_{\rm BH}$ for \src is inferred using the BLR velocity dispersion
which is {$10^{7.29}$}\msun~(Yan19).  
The rise and decline time-scale are defined as the duration it takes for a TDE 
to rise from half-peak to peak and to decline from peak to half-peak, respectively. 
}
\label{fig:compare_TDE}
}
\end{figure}

{In the scenario of a TDE, both the rise and decline time-scales of \src appears notably longer compared to other optical TDEs in the ZTF sample (Figure \ref{fig:compare_TDE}). 
By characterizing the light curve evolutionary speed with the duration above half-peak \citep{Yao2023}, 
we found the rise time-scale of \srcs's optical outburst is $t_{\rm 1/2, rise}\sim 61$ 
days, and the decline time-scale $t_{\rm 1/2, decline}\sim 228$  
days, making it an unusually slow-evolved TDE candidate. 
The slower evolution of \src  could be explained by 
its larger black hole mass as well as the larger mass and radius of the 
disrupted star (Section 3.1), 
as the luminosity of TDE is expected to follow the fallback rate 
of the stellar debris with a relationship \citep{Rees1988}: 
\begin{equation*}
   t_{\rm fb}=41 M^{1/2}_{6}r_{*}^{3/2}m_{*}^{-1}\beta^{-3} {\rm days}
\end{equation*}
where $M_{6}$ is the BH mass in $10^{6}$\msun, $r_{*}$ and $m_{*}$ are the star's radius and mass in $R_\odot$ and $M_\odot$, 
and $\beta=R_{\rm T}/R_{\rm p}$ is the penetration factor \citep{Gezari2017, Bonnerot2020}. 
In addition, there could be slowed disk formation and inefficient circularization, 
as a relatively large viscous time-scale ($T_{\rm viscous}= 30.5_{-19.7}^{+16.4}$ days) 
was inferred from 
the {\tt MOSFiT} fits to the UV-optical light curves. 
In this case, the luminosity evolution can be further slowed, because the viscous delay can 
flatten the peak of the light curve \citep{Mockler2019}. 
}

Recent optical time-domain surveys have unveiled further complexities in the declining light curves of TDEs \citep[][]{Yao2023}. 
For instance, the UV/optical light curves of the TDE AT2018fyk \citep{Wevers2019}, AT2020acka and AT2021uqv \citep{Yao2023} show a plateau phase of 130--150 days between an initial and second monotonic decline. 
This is very similar to that seen in \srcs, though on shorter timescales. 
In the scenario of a TDE, the plateaued UV/optical emission might be powered by 
the reprocessing of the accretion luminosity, indicating {a delayed formation of a compact accretion disk}. 
In addition to broad Balmer and He I lines, narrow emission lines also emerge in the optical
spectra of \src during the plateau phase \citep{Yan2019}, which is in line with the reprocessing scenario. 
However, we did not observe the brightening in the X-ray emission at later times when 
the UV/optical light curve declines again after the plateau. 
This suggests that the accretion disk might have evolved too fast or the X-ray emission might be too faint to be detected. 



\subsection{On the Origin of the Delayed Radio Flare}

Using an equipartition analysis, we found that the energy and the velocity of the outflow that powers the
radio flare are similar to other non-relativistic TDEs (Figure \ref{fig:param_compare}). 
{However, the timescale 
and the steep decline ($F_{\rm \nu}\propto t^{-6.36}$) 
are unusual compared}
to non-relativistic TDEs with “prompt” emission (Figure \ref{fig:radio_lc} and Section 3.2). 
The evolution of equipartition radius with the time indicates that the outflow 
was launched with a substantial delay of $\sim$1150 days relative to the time of optical discovery, 
regardless of the outflow being spherical or collimated. 
In this case, the post-peak radio flux decline can be described by a power-law of $t^{-2.53}$, which is not in tension with a decelerating outflow interacting with the CNM. 
Therefore, \src represents one of few TDEs with long delayed radio emission \citep[$t_{\rm occur}>1000$ days,][]{Cendes2024}, 
the nature of which is still in dispute. 

We first explore the possibility that the late radio emission in \src is due to a relativistic jet with an off-axis viewing orientation. 
In this model, the off-axis jet was initially launched at the time of TDE, 
which remains collimated with the radio-emitting area increasing over time \citep{Matsumoto2023, Sfaradi2024}. 
The evolution of radio emitting region eventually intersects the light of sight to the observer, 
resulting in a delayed radio flare. 
By generalizing the conventional equipartition method for an arbitrary viewing angle, 
\citet{Matsumoto2023} proposed that if the radio-emitting source is initially an off-axis jet and evolves into the on-axis Newtonian branch, 
its apparent velocity $\beta_{\rm eq, N}$ should increase to across the critical value of 
$\beta_{\rm eq, N}\simgt0.23$. 
Such a model is potentially relevant for TDEs in which the radio emission is still rising, such as AT2018hyz \citep{Matsumoto2023, Sfaradi2024}. 
However, the radio emission of \src has already peaked, and the nearly linear increase in $R_{\rm eq}$ 
implies $\beta_{\rm eq, N}\sim0.05-0.11$, depending on the outflow geometry assumed. 
On a basis of the best-fit synchrotron model to the radio SED evolution (Figure \ref{fig:radio_req}, left), 
we find both the peak flux densities and frequencies are decreasing over time, 
with $F_{\rm peak}\propto t^{-1.79}$ and $\nu_{\rm peak}\propto t^{-1.89}$. 
Following the formalism of \citet{Matsumoto2023}, the apparent velocity of the radio emitting source 
is found to evolve with $\beta_{\rm eq, N}\propto t^{-0.8}$.  
Therefore, $\beta_{\rm eq, N}$ will continue to decrease and do not cross the threshold\footnote{Note that in the realistic case of the jet symmetry (e.g., $\theta<\pi/2$), the critical 
value becomes larger $\beta_{\rm eq, N}\simeq0.44$ \citep{Beniamini2023}. 
In this case, it is more unlikely that the transition to the
Newtonian branch will happen. }
of $\beta_{\rm eq, N}\simgt0.23$, 
disfavoring the off-axis jet that launched at the time of optical discovery as the origin of delayed radio flare in \srcs. 

The combined radio flux and spectral evolution properties may point to a radio-emitting process that 
occurs at a later time, possibly driven by the interaction of a delayed ejection of an outflow (either 
relativistic or sub-relativistic) with the CNM. 
One possibility is that the onset of accretion onto the SMBH is delayed due to 
the long timescales for debris circularization, 
leading to the late-time launching of outflows. 
However, such a scenario seems disfavored as the flattening of UV/optical emission has been 
observed in the luminosity evolution (Figure \ref{fig:opt_lc}, see also Yan19), 
which could be explained by the reprocessing of accretion luminosity. 
This suggests that 
the accretion disk might have formed effectively, around 200 days after the optical peak, 
hence is inconsistent with the timescale of outflow launching we inferred for \src from radio observations. 
X-ray observations are crucial in constraining the onset of accretion, 
but no transient X-ray emission is observed (Section 2.2). 
A scenario in which an outflow travels in an inhomogeneous CNM 
whose density profile flattens outside the Bondi radius \citep{Matsumoto2024}, 
producing a late-time radio flare, is also unlikely. 
In this latter case, the radio flux density is expected to decrease slowly after peak ($\nu F_{\nu}\propto t^{-1}$), 
in contrast to what we observed (Figure \ref{fig:radio_sed}), 
even if considering that the outflow launching is itself delayed by several hundred days. 

Recently, \citet{Teboul2023} proposed a unified model for jet production in TDEs \citep[see also][]{Lu2024} 
in which the jet might be present during the entire period of the TDE evolution,  
but it is initially choked by the disk-wind ejecta.  
In the model, the jet will be precessing to align with the SMBH spin axis. 
Relatively weak jets align through the hydrodynamic mechanism,
which requires longer time (several months or years) to escape from the disk-wind cocoon and expand into the CNM, 
leading to the delayed radio emission. 
Such a delayed escaping jet can possess a characteristic energy
up to $E\sim10^{50}$ erg, comparable to that observed in \srcs. 
However, the model assumes that the black hole is accreting at a highly super-Eddington rate, 
producing a quasi-spherical mass-loaded disk-wind envelope to confine the jet. 
This is inconsistent with the inferred sub-Eddington accretion rate of \src ($\sim$$0.051\dot{M}$$_{\rm Edd}$) at the peak of the outburst. 
Alternatively, the delayed radio emission can be produced by the delayed ejection of an outflow, 
perhaps associated with the state transition in the SMBH accretion.  
This is predicted to occur in X-ray binaries \cite[XRBs,][]{Fender2004} when the mass accretion rate decreases to across a critical 
threshold, typically a few percent of the Eddington luminosity \citep{Done2007}. 
At the outflow launching time ($t\sim1150$ days), 
{we find that both the optical and MIR emission of \src has decreased by a factor of $\sim$ 2,}
lower than that in the plateau phase. 
This corresponds to the decrease in the mass accretion rate from $\sim$$0.022\dot{M}$$_{\rm Edd}$ 
to {$\sim$ 0.012$\dot{M}$$_{\rm Edd}$}, which is not at odds 
with the possibility of a state change if in analogy with XRBs. 
Note that a similar scenario has been proposed to explain the late-time radio brightening 
in the TDE ASASSN-15oi \citep{Horesh2021a}, though the details of what triggers the state transition remain unclear. 



\section{Conclusion}

\src was reported to be a candidate ``turn-on" AGN in a LINER galaxy, for which the post-outburst UV/optical 
emission exhibits an unusual plateau phase. 
We present new results from the analysis of the complete multi-wavelength dataset 
spanning $\sim$3000 days since its discovery. 
We find that the UV/optical and MIR emission have faded back to the baseline level, 
suggesting that the outbursts are likely powered by a short-lived accretion event such as a TDE. 
{In this case, the long rise and decline time-scales make \src an unusually slow-evolved optical TDE.}
In addition, we find a delayed radio brightening followed by a steep flux decay, 
with a radio luminosity as high as $\nu L_{\nu}(\rm 5.5~GHz)$ $\sim 1.9 \times 10^{39}$ \erg.  
The radio emitting region is compact with a size of $<$1.58 pc, at least at $t=1946$ days since discovery, 
and there is no resolved component at a milliarcsecond scale.
Using an equipartition analysis, the radio flux and SED evolution can be described by an 
outflow interacting with CNM, which was launched at $t\approx$1150 days after optical discovery 
with a velocity of $\beta \simlt$0.1. 
The outflow kinetic energy is $\simgt10^{50}$ erg, which is in excess of most previous non-relativistic TDEs. 
The CNM density profile can be described as $R^{-2.5}$, typical of previous TDEs, 
with a potential further steepening at a larger radius.   
The observed rise in radio flux coupled with the disappearing plateau in the UV/optical light curves 
points to the scenario involving a delayed ejection of an outflow, perhaps from a state transition in the disk. 
To better study the detailed process of outflow launching, late-time radio observing campaigns of a larger sample of optical TDEs, especially those with plateau emission, are encouraged. 

\acknowledgments{
The data presented in this paper are based on observations
made with the Karl G. Jansky Very Large Array from the program VLA/18B-086, VLA/20A-251, VLA/21A-146 and VLA/21B-168) 
the Very Long Baseline Array from the project VLBA/21A-045, the Giant Metrewave Radio Telescope from the project 37\_133 and 41\_065, and the MeerKAT from the project SCI-20230907-XS-01.
We thank the staff of the VLA, VLBA, GMRT and MeerKAT that made these observations possible. 
The National Radio Astronomy Observatory is a facility of the
National Science Foundation operated under cooperative agreement
by Associated Universities, Inc.  
GMRT is run by the National Centre for Radio Astrophysics of the Tata Institute of Fundamental Research. 
The MeerKAT telescope is operated by the South African Radio Astronomy Observatory, which is a facility of the National Research Foundation, an agency of the Department of Science and Innovation.
We acknowledge the use of the ilifu cloud computing facility (\url{www.ilifu.ac.za}) and the Inter-University Institute for Data Intensive Astronomy (IDIA), a partnership of the University of Cape Town, the University of Pretoria and the University of the Western Cape. 
 This work has made use of the “MPIfR S-band receiver system” designed, constructed and maintained by funding of the MPI für Radioastronomy and the Max-Planck-Society. 
 The work is supported by the SKA Fast Radio Burst
and High-Energy Transients Project (2022SKA0130102), and the National Science Foundation of China (NSFC) through grant No. 12192220, 12192221 and 11988101. 
X.S. acknowledges the science research grants from the China
Manned Space Project with NO. CMSCSST-2021-A06.  
Y.C. thanks the Center for Astronomical Mega-Science, Chinese Academy of Sciences, for the FAST distinguished young researcher fellowship (19-FAST-02). Y.C. also acknowledges the support from the National Natural Science Foundation of China (NSFC) under grant No. 12050410259 and the Ministry of Science and Technology (MOST) of China grant no. QNJ2021061003L. 
F.X.A acknowledges the support from the National Natural Science Foundation of China (12303016) and the Natural Science Foundation of Jiangsu Province (BK20242115). 
D.L. is a New Cornerstone Investigator.
 }
 



\software{CASA \citep[v5.3.0 and v5.6.1; ][]{McMullin2007}, 
MOSFiT \citep{Guillochon2018}, Astropy \citep{Astropy2013, Astropy2018, Astropy2022}, 
AIPS \citep{Greisen2003}, DiFX software correlator \citep{Deller2011}, DIFMAP \citep{Shepherd1997}.
 }

\bibliographystyle{aasjournal}
\bibliography{ms_j1115_radio.bib}

\end{document}